\documentclass[secnumarabic,aps,pra,amsfonts,twocolumn,%
showkeys
]{revtex4-1}

\renewcommand{\bibnumfmt}[1]{\mbox{#1}. }
\usepackage{bm,amsfonts,amsmath,amsthm}
\usepackage{placeins}
\usepackage{longtable}
\usepackage{graphicx}
\newcommand{\gl}[1]{(\ref{#1})}

\begin{document}

\title{Constraining Forces Stabilizing Superconductivity in Bismuth}
\author{Ekkehard Kr\"uger} 
\affiliation{Institut f\"ur Materialwissenschaft, Materialphysik,
  Universit\"at Stuttgart, D-70569 Stuttgart, Germany}
%
\date{\today}
\begin{abstract}
 As shown in former papers, the nonadiabatic Heisenberg
  model presents a novel mechanism of Cooper pair formation generated
  by the strongly correlated atomic-like motion of the electrons in
  narrow, roughly half-filled ``superconducting bands''.  These are
  energy bands represented by optimally localized spin-dependent
  Wannier functions adapted to the symmetry of the material under
  consideration. The formation of Cooper pairs is not the result of an
  attractive electron-electron interaction but can be described in
  terms of quantum mechanical constraining forces constraining the
  electrons to form Cooper pairs. There is theoretical and
  experimental evidence that only this nonadiabatic mechanism
  operating in superconducting bands may produce {\em eigenstates} in
  which the electrons form Cooper pairs. These constraining forces
  stabilize the Cooper pairs in any superconductor, whether
  conventional or unconventional. Here we report evidence that also
  the experimentally found superconducting state in bismuth at ambient
  as well as at high pressure is connected with a narrow, roughly
  half-filled superconducting band in the respective band structure.
  This observation corroborates once more the significance of
  constraining forces in the theory of superconductivity.
\end{abstract}

\keywords{superconductivity; bismuth at ambient pressure; Bi--I; bismuth at
  high pressure; Bi--V; constraining forces; nonadiabatic Heisenberg model}
\maketitle

\section{Introduction}
\label{sec:introduction}
Bismuth shows sequential structure transition as function of the
applied pressure, as summarized in an illustrative form by O. Degtyareva {\em et
  al.}~\cite{degty}:
\begin{widetext}
\begin{equation}
  \label{eq:1}
  \text{Bi--I}\ \xrightarrow{\text{2.55 GPa}}\ \text{Bi--II}\
  \xrightarrow{\text{2.7 GPa}}\ \text{Bi--III}\ \xrightarrow{\text{7.7 GPa}}\ \text{Bi--V}\ < \text{122 GPa} 
\end{equation}
\end{widetext}
At ambient pressure, Bi crystallizes in the structure Bi--I, an
As-type structure with a trigonal (rhombohedral) space group and two
atoms in the unit cell~\cite{donohue}. This structure is stable up to
a pressure of 2.55 GPa. Then, with increasing pressure, Bi undergoes
the monoclinic structure Bi--II and the host-guest structure
Bi--III. A further structure called Bi--IV exists above the
temperature of 450 K and is not relevant in this paper. Between a
pressure of 7.7 and (at least) 122 GPa, the cubic Bi--V phase is
stable~\cite{degty}.

It is interesting, that all these Bi phases become superconducting at
low temperatures. The Bi--I phase is superconducting with the
extremely low transition temperature $T_c = 0.53$mK~\cite{prakash}. In
the Bi--II and Bi--III structures, the transition temperature
increases with increasing pressure from about 4 K to 7 K. Finally, in
the Bi--V phase, $T_c$ has the maximum value of about 8
K~\cite{li}. The different values of $T_c$ are evidently connected
with the different crystal structures since $T_c$ changes
discontinuously at the transitions from one structure to
another~\cite{li}.

This striking symmetry-dependence of the superconducting transition
temperature suggests that also in bismuth superconductivity is
connected with narrow, roughly half-filled ``superconducting
bands''. A closed energy band (Definition 2 of Ref.~\cite{theoriewf})
with optimally localized, symmetry-adapted, and spin-dependent Wannier
functions is called superconducting band (Definition 22 of
Ref.~\cite{theoriewf}) because those metals (and only those metals)
that possess such a narrow, roughly half-filled superconducting band
in its band structure experimentally prove to be superconductors, see
the Introduction of Ref.~\cite{theoriewf}. This observation can be
interpreted within the group-theoretical nonadiabatic Heisenberg model
(NHM)~\cite{enhm}, a new model of strongly correlated atomic-like
electrons. Within this model, the formation of Cooper pairs is still
mediated by boson excitations (responsible, as usual, for the isotope
effect). However, these boson excitations produce constraining forces
as they are familiar from classical mechanics: below $T_c$, they
reduce the degrees of freedom of the electron system by forcing the
electrons to form Cooper pairs. A short description of the NHM and
this novel mechanism of Cooper pair formation is given in Secs.~2
and~3, respectively, of Ref.~\cite{josybacuo7}. In
Sec.~\ref{sec:discussion} we shall summarize this new concept of
superconductivity in the form of single statements.

There is theoretical evidence that the constraining forces
operating in narrow, roughly half-filled superconducting bands are
required for the Hamiltonian of the system to possess {\em
  eigenstates} in which the electrons form Cooper pairs~\cite{josn}.
The aim of the present paper is to corroborate this important
assertion by showing that also the experimentally established
superconductivity in bismuth~\cite{prakash,li} is evidently connected
with superconducting bands.



\begin{figure*}[t]
\centering
\includegraphics[width=\textwidth]{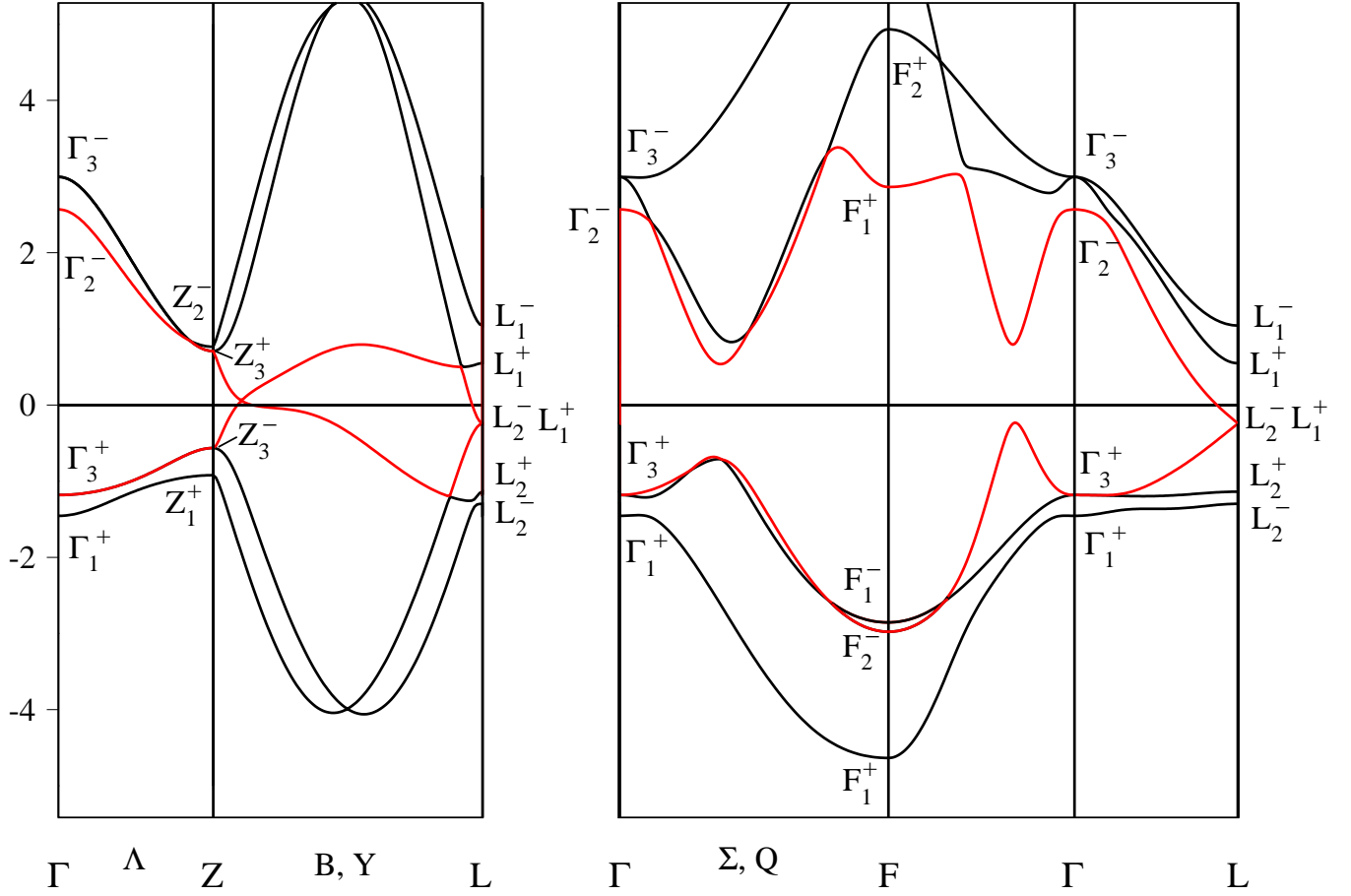}
\caption{ Band structure of Bi--I calculated by the FHI-aims program
  \protect\cite{blum1,blum2}, using the structure parameters given by
  O. Degtyareva {\em et al.}~\cite{degty}. The symmetry labels are
  determined by the author.  Bi--I has the trigonal space group
  $R\overline{3}m$~\cite{donohue} (international number 166), the
  notations of the points and lines of symmetry in the Brillouin zone
  for $\Gamma_{rh}$ follow Fig. 3.11 (b) of Ref.~\cite{bc}, and the
  symmetry labels are defined in Table~\ref{tab:rep166}. $E_F$ denotes
  the Fermi level. The band highlighted in red is the superconducting
  band.}
\label{fig:bs_166}
\end{figure*}   


In this context, we consider (in the following
Sec.~\ref{sec:supbands}) only the two structures Bi--I and Bi--V at
the beginning and the end of the sequence~\gl{eq:1}. Bi--I and Bi--V
possess the lowest and highest superconducting transition
temperatures, respectively. Bi--II is not very informative within the
NHM since it has only a low monoclinic symmetry. At this stage, it
would be complicated to apply the NHM to the incommensurate host-guest
structure of Bi--III.  Both Bi--I and Bi--V, on the other hand, have
clear symmetries with the trigonal space group $R\overline{3}m$ (166)
and the cubic space group $Im3m$ (229),
respectively~\cite{donohue,degty}. Bi--V even has the highest possible
symmetry in a solid state with allows the NHM to make clear
predictions.

\section{Superconducting bands in the band structure of bismuth}
\label{sec:supbands}
\subsection{Band structure of Bi--I}
The band structure of Bi--I is depicted in Fig.~\ref{fig:bs_166}. The
Bloch functions of the band highlighted in red are labeled by the
single-valued representations
\begin{equation}
\label{eq:4}
\begin{array}{llllllllll}
\Gamma^-_2,& \Gamma^+_3; \ &
Z^+_3,& Z^-_3;\ &L^+_1,& L^-_2;\ & F^+_1,& F^-_2.\\  
\end{array}
\end{equation}

It is clear that this band (or any other band in the band structure)
does not contain a closed band (Definition 2 of Ref.~\cite{theoriewf})
with the symmetry of band 1 or band 2 in Table~\ref{tab:wf166},
meaning that we cannot unitarily transform the Bloch functions into
best localized and symmetry-adapted Wannier functions situated at the
Bi atoms. The situation is changed when we consider the double-valued
representations of the Bloch functions:



\begin{figure*}[t]
\centering
\includegraphics[width=\textwidth]{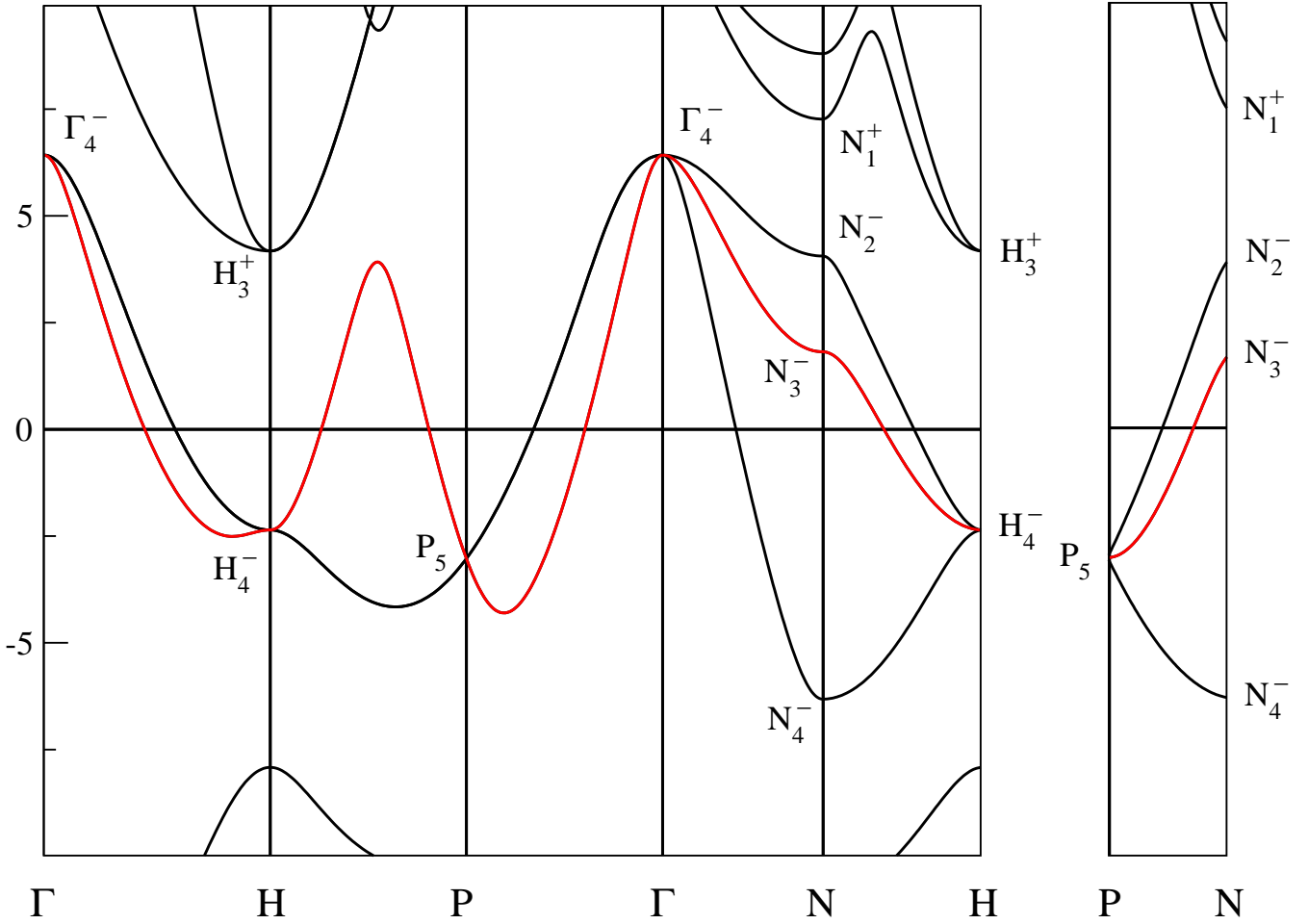}
\caption{Band structure of Bi--V at the pressure of 13.5 GPa
  calculated by the FHI-aims program \protect\cite{blum1,blum2}, using
  the structure parameters at this pressure as given by O. Degtyareva
  {\em et al.}~\cite{degty}. The symmetry labels are determined by the
  author. Bi--V has the cubic space group $Im3m$~\cite{degty}
  (international number 229), the notations of the points and lines of
  symmetry in the Brillouin zone for $\Gamma^v_{c}$ follow Fig. 3.15
  of Ref.~\cite{bc}, and the symmetry labels are defined in
  Table~\ref{tab:rep229}. $E_F$ denotes the Fermi level. The band
  highlighted in red forms the superconducting band.}
\label{fig:bs_229}
\end{figure*}

According to Table~\ref{tab:falten166}, we may unitarily transform the
Bloch functions~\gl{eq:4} into Bloch functions labeled by the
double-valued representations,
\begin{equation}
  \label{eq:3}
\begin{array}{lclp{.3cm}lcl}
\Gamma^-_2 & \rightarrow & \underline{\Gamma^-_4}, && 
\Gamma^+_3 & \rightarrow & \underline{\Gamma^+_4} + \Gamma^+_5 + \Gamma^+_6 ;\\[.2cm] 
Z^+_3 & \rightarrow & \underline{Z^+_4} + Z^+_5 + Z^+_6 ,&&  
Z^-_3 & \rightarrow & \underline{Z^-_4} + Z^-_5 + Z^-_6;\\[.2cm]  
L^+_1 & \rightarrow & \underline{L^+_3} + \underline{L^+_4},&&  
L^-_2 & \rightarrow & \underline{L^-_3} + \underline{L^-_4};\\[.2cm]  
F^+_1 & \rightarrow & \underline{F^+_3} + \underline{F^+_4},&&  
F^-_2 & \rightarrow & \underline{F^-_3} + \underline{F^-_4}.\\[.2cm]  
\end{array}
\end{equation}
The underlined representations belong to the band listed in
Table~\ref{tab:wf166Z}. Thus,
we can unitarily transform the Bloch functions of this band
into {\em spin-dependent} Wannier functions being best
localized, centered at the Bi atoms, and symmetry-adapted to the group
$R\overline{3}m$. Consequently, according to Definition 22 of
Ref.~\cite{theoriewf}, the band highlighted in red is a
superconducting band. 

\subsection{Band structure of Bi--V}
The band structure of Bi--V is depicted in Fig.~\ref{fig:bs_229}. The
Bloch functions of the band highlighted in red now are labeled by the
single-valued representations
\begin{equation}
\label{eq:5}
\begin{array}{llll}
\Gamma^-_4;\ & H^-_4;\ & P_5;\ & N^-_3.\\  
\end{array}
\end{equation}

Again, this band (or any other band in the band structure) does not
contain a closed band (Definition 2 of Ref.~\cite{theoriewf}) with the
symmetry of the bands listed in Table~\ref{tab:wf229}. Hence, we
cannot unitarily transform the Bloch functions into best localized and
symmetry-adapted Wannier functions situated at the Bi atoms.
According to Table~\ref{tab:falten229}, we may unitarily transform the
Bloch functions~\gl{eq:5} into Bloch functions labeled by the
double-valued representations,
\begin{equation}
  \label{eq:6}
\begin{array}{lclp{.3cm}lcl}
\Gamma^-_4 & \rightarrow & \underline{\Gamma^-_6} + \Gamma^-_8,\\ 
 H^-_4 & \rightarrow & \underline{H^-_6} + H^-_8,\\  
 P_5 & \rightarrow & \underline{P_7} + P_8,\\  
 N^-_3 & \rightarrow & \underline{N^-_5}.\\  
\end{array}
\end{equation}
The underlined representations belong to band 4 listed in
Table~\ref{tab:wf229Z}. Thus,
we can unitarily transform the Bloch functions of this band
into {\em spin-dependent} Wannier functions being best
localized, centered at the Bi atoms, and symmetry-adapted to the group
$Im3m$. Consequently, according to Definition 22 of
Ref.~\cite{theoriewf}, the band highlighted in red is a
superconducting band.

\subsection{Interpretation}
Both structures Bi--I and Bi--V possess a superconducting band in their
band structure that
\begin{itemize}
\item is one of the narrowest bands in the band structure;
\item is nearly half filled;
\item and comprises a great part of the electrons at the Fermi level.
\end{itemize}
Consequently, the NHM predicts that both phases become superconducting
below a transition temperature $T_c$.  

The superconducting band of Bi--I (Fig.~\ref{fig:bs_166}) even
comprises all the electrons at the Fermi level. However, the small
Fermi surface and the small density of states at the Fermi level
results in the extremely low superconducting transition temperature of
$T_c = 0.53$mK~\cite{prakash}.

The superconducting band of Bi--V (Fig.~\ref{fig:bs_229}) closely
resembles the superconducting band of niobium as depicted, e.g., in
Fig.~1 of Ref.~\cite{josn}: both nearly half-filled bands have
comparable widths and comprise a comparable part of the Fermi
level. Consequently, we may expect that both the Bi--V phase of
bismuth and the elemental metal niobium have similar transition
temperatures. Indeed, we have $T_c \approx 8$K and $T_c = 9.2$K for
Bi--V and niobium, respectively. Narrow and half-filled
superconducting bands rarely arise in crystals with the high bcc
symmetry. So the elemental bcc metals Ta, W, and Mo possess
superconducting bands which are far from being half-filled and,
consequently, have lower transition temperatures. In the band
structures of the most elemental metals (such as Li, Na, K, Rb, Cs, Ca
Cu, Ag, and Au), narrow, roughly half-filled superconducting bands
cannot be found and, hence, these metals do not become
superconducting~\cite{es2}. Consequently, there is high evidence that
the superconducting state in Bi--V is connected with the narrow and
almost perfectly half-filled superconducting band in the band
structure of this phase.

\section{Results}
\label{sec:results}
In terms of superconducting bands, the NHM confirms the experimental
observations that
\begin{itemize}
\item the Bi--I phase
(i.e., bismuth at ambient pressure) becomes superconducting below an extremely
low transition temperature and
\item the Bi--V phase (i.e., bismuth at high
pressure) becomes superconducting below a transition temperature
comparable with the transition temperature of niobium.
\end{itemize} 

\section{Discussion}
\label{sec:discussion}
This group-theoretical result demonstrates again~\cite{theoriewf} the
significance of the theory of superconductivity defined within the
NHM. We summarize the main features of this novel concept of
superconductivity (a more detailed description is given in
Ref.~\cite{josybacuo7}):

\begin{itemize}
\item The NHM is based on three postulates~\cite{enhm} concerning the
  {\em atomic-like} motion of the electrons in narrow, half-filled
  energy bands as it was already considered by Mott \cite{mott} and
  Hubbard\cite{hubbard}.
\item The postulates of the NHM are physically evident and require the
  introduction of {\em nonadiabatic} localized states of well-defined
  symmetry emphasizing the {\em correlated} nature of any atomic-like
  motion.
\item The atomic-like motion is determined by the conservation of the
  total crystal-spin angular momentum which must be satisfied in the
  nonadiabatic system. In a narrow, roughly half-filled
  superconducting band this conservation law plays a crucial role
  because the localized (Wannier) states are spin-dependent.
\item The strongly correlated atomic-like motion in a narrow, roughly
  half-filled superconducting band produces an interaction between the
  electron spins and ``crystal-spin-1 bosons'': at any electronic
  scattering process two crystal-spin-1 bosons are excited or absorbed
  in order that the total crystal-spin angular momentum stays
  conserved.
\item Crystal-spin-1 bosons are the {\em energetically lowest}
  localized boson excitations of the crystal that possess the
  crystal-spin angular momentum $1\cdot\hbar$ and are sufficiently
  stable to transport it (as Bloch waves) through the crystal.
\item The spin-boson interaction in a narrow, roughly half-filled
  superconducting band leads to the formation of Cooper pairs below a
  transition temperature $T_c$.
\item The Cooper pairs arise inevitably since any electron state in
  which the electrons possess their full degrees of freedom violates
  the conservation of crystal-spin angular momentum. 
\item This influence of the crystal-spin angular momentum may be
  described in terms of constraining forces that constrain the
  electrons to form Cooper pairs. This feature distinguishes the
  present concept from the standard theory of superconductivity.
\item As already mentioned in Sec.~\ref{sec:introduction}, there is
  evidence that {\em only} these constraining forces may produce
  superconducting {\em eigenstates}.
\item Hence, the constraining forces are responsible for all types of
  superconductivity, i.e., conventional, high-$T_c$ and other
  superconductivity.
\item Crystal-spin-1 bosons are coupled phonon-plasmon modes that
  determine the type of the superconductor.
\item In the isotropic lattices of the transition elements,
  crystal-spin-1 bosons have dominant phonon character and confirm the
  electron-phonon mechanism that enters the BCS theory~\cite{bcs} in
  these materials.
\item Phonon-like excitations are not able to transport crystal-spin
  angular-momenta within the anisotropic materials of the high-$T_c$
  superconductors~\cite{ehtc}, often containing two-dimensional
  layers. Within these anisotropic materials, the crystal-spin-1
  bosons are energetically higher lying excitations of dominant
  plasmon character leading to higher superconducting transition
  temperatures~\cite{bcs}.
\item The theory of superconductivity as developed so far is valid
  without any restrictions in narrow, roughly half-filled
  superconducting bands because constraining forces do not alter the
  energy of the electron system.
\item However, the standard theory may furnish inaccurate information
  if no narrow, roughly half-filled superconducting band exists in the
  band structure of the material under consideration.
\end{itemize}
It is clear that this concept of superconductivity as developed in the
last 40 years should be further refined in the future.

\acknowledgments{I am very indebted to Guido Schmitz for his support
  of my work. I thank G\"unter Zerweck for his valuable reference to bismuth.}

\appendix
\onecolumngrid
\section{Group-theoretical tables for the trigonal space group $R\overline{3}m$ (166)
of Bi--I }

It is sometimes useful to represent trigonal (rhombohedral) systems in
a hexagonal coordinate system. In this case, the unit cell contains
two additional inner points which, however, are connected to each
other and to the points at the corners by the translation symmetry of
the system. In the framework of the group theory of Wannier functions
as presented in Ref.~\cite{theoriewf}, the inner points of a unit cell
must not be connected by the translation symmetry. Thus, the group
theory of Wannier functions is not applicable to trigonal system
represented by hexagonal axes. Therefore, in the present paper, we use
exclusively the trigonal coordinate system as given in Table 3.1 of
Ref.~\cite{bc}.


\begin{table}
\caption{
  Character tables of the single-valued irreducible representations of the
  trigonal space group $R\overline{3}m =
  \Gamma_{rh}D^{5}_{3d}$ (166) of 
  Bi--I. 
  \label{tab:rep166}}
\begin{tabular}[t]{ccccccc}
\multicolumn{7}{c}{$\Gamma (000)$, $Z (\frac{1}{2}\frac{1}{2}\frac{1}{2})$}\\
 & $E$ & $I$ & $S^{\pm}_6$ & $C^{\pm}_3$ & $C'_{2i}$ & $\sigma_{di}$\\
\hline
$\Gamma^+_1$, $Z^+_1$ & 1 & 1 & 1 & 1 & 1 & 1\\
$\Gamma^+_2$, $Z^+_2$ & 1 & 1 & 1 & 1 & -1 & -1\\
$\Gamma^-_1$, $Z^-_1$ & 1 & -1 & -1 & 1 & 1 & -1\\
$\Gamma^-_2$, $Z^-_2$ & 1 & -1 & -1 & 1 & -1 & 1\\
$\Gamma^+_3$, $Z^+_3$ & 2 & 2 & -1 & -1 & 0 & 0\\
$\Gamma^-_3$, $Z^-_3$ & 2 & -2 & 1 & -1 & 0 & 0\\
\hline\\
\end{tabular}\hspace{.5cm}
\begin{tabular}[t]{ccccc}
\multicolumn{5}{c}{$L (0\frac{1}{2}0)$}\\
 & $E$ & $C'_{22}$ & $I$ & $\sigma_{d2}$\\
\hline
$L^+_1$ & 1 & 1 & 1 & 1\\
$L^-_1$ & 1 & 1 & -1 & -1\\
$L^+_2$ & 1 & -1 & 1 & -1\\
$L^-_2$ & 1 & -1 & -1 & 1\\
\hline\\
\end{tabular}\hspace{.5cm}
\begin{tabular}[t]{ccccc}
\multicolumn{5}{c}{$F (\frac{1}{2}\frac{1}{2}0)$}\\
 & $E$ & $C'_{23}$ & $I$ & $\sigma_{d3}$\\
\hline
$F^+_1$ & 1 & 1 & 1 & 1\\
$F^-_1$ & 1 & 1 & -1 & -1\\
$F^+_2$ & 1 & -1 & 1 & -1\\
$F^-_2$ & 1 & -1 & -1 & 1\\
\hline\\
\end{tabular}
\ \\
\begin{flushleft}
Notes to Table~\ref{tab:rep166}
\end{flushleft}
\begin{enumerate}
\item $i = 1,2,3.$
\item The symmetry elements are labeled in the Sch\"onflies notation
  as illustrated, e.g., in Table 1.2 of Ref.~\cite{bc}.  
\item The character tables are determined from Table 5.7 of
  Ref.~\protect\cite{bc}.
\item The notations of the points of symmetry follow Fig. 3.11 (b) of 
Ref.~\cite{bc}. 
\end{enumerate}
\end{table}




\begin{table}
\caption{
  Character tables of the single-valued irreducible representations of
  the point 
  group $C_{3v}$ of the positions of the Bi atoms (Definitions 11 and 12 of Ref.~\cite{theoriewf})
in Bi--I.
\label{tab:repBi}}
\centering
\begin{tabular}[t]{cccc}
 & $E$ & $C^{\pm}_3$ & $\sigma_{di}$\\
\hline
$\bm{d}_{1}$ & 1 & 1 & 1\\
$\bm{d}_{2}$ & 1 & 1 & -1\\
$\bm{d}_{3}$ & 2 & -1 & 0\\
\hline\\
\end{tabular}\hspace{1cm}

$i = 1,2,3.$
\end{table}




\begin{table}
\caption{
    Compatibility relations between the single-valued (upper row) and
    double-valued (lower row) representations of the space group
    $R\overline{3}m$. 
\label{tab:falten166}}
\centering
\begin{tabular}[t]{cccccc}
\multicolumn{6}{c}{$\Gamma (000)$, $Z (\frac{1}{2}\frac{1}{2}\frac{1}{2})$}\\
\hline
$R^+_1$ & $R^+_2$ & $R^-_1$ & $R^-_2$ & $R^+_3$ & $R^-_3$\\
$R^+_4$ & $R^+_4$ & $R^-_4$ & $R^-_4$ & $R^+_5$ + $R^+_6$ + $R^+_4$ & $R^-_5$ + $R^-_6$ + $R^-_4$\\
\hline\\
\end{tabular}\hspace{1cm}
\begin{tabular}[t]{cccc}
\multicolumn{4}{c}{$L(0\frac{1}{2}0)$, $F(\frac{1}{2}\frac{1}{2}0)$}\\
\hline
$R^+_1$ & $R^-_1$ & $R^+_2$ & $R^-_2$\\
$R^+_3$ + $R^+_4$ & $R^-_3$ + $R^-_4$ & $R^+_3$ + $R^+_4$ & $R^-_3$ + $R^-_4$\\
\hline\\
\end{tabular}
\ \\
\begin{flushleft}
Notes to Table~\ref{tab:falten166}
\end{flushleft}
\begin{enumerate}
\item The letter $R$ stands for the letter denoting the relevant point of
  symmetry.  For example, at point $F$ the representations $R^+_1, R^+_2, \ldots$
  stand for $F^+_1, F^+_2, \ldots$ .
\item Each column lists the double-valued representation $R_i\times
  {\bm d}_{1/2}$
  below the single-valued representation $R_i$, where ${\bm d}_{1/2}$ denotes the
  two-dimensional double-valued representation of the three-dimensional
  rotation group $O(3)$ given, e.g., in Table 6.1 of Ref.~\cite{bc}.
\item The single-valued representations are defined in
  Table~\ref{tab:rep166}. 
\item The notations of double-valued representations follow strictly
  Table 6.13 (and Table 6.14) of Ref.~\cite{bc}. In this paper the
  double-valued representations are not explicitly given but are
  sufficiently defined by this table.
\end{enumerate}
\end{table}
 

\begin{table}
\caption{
Single-valued representations of all the energy bands in the space
group $R\overline{3}m$ of 
Bi--I with symmetry-adapted and optimally  
localized usual (i.e., spin-independent) Wannier functions centered at the Bi atoms. 
\label{tab:wf166}}
\centering
\begin{tabular}[t]{cccccccc}
& Bi($zzz$) & Bi($\overline{z}\overline{z}\overline{z}$) & $K$ & $\Gamma$ & $Z$ & $L$ & $F$\\
\hline
Band 1 & $\bm{d}_{1}$ & $\bm{d}_{1}$ & OK & $\Gamma^+_1$ + $\Gamma^-_2$ & $Z^+_1$ + $Z^-_2$ & $L^+_1$ + $L^-_2$ & $F^+_1$ + $F^-_2$\\
Band 2 & $\bm{d}_{2}$ & $\bm{d}_{2}$ & OK & $\Gamma^+_2$ + $\Gamma^-_1$ & $Z^+_2$ + $Z^-_1$ & $L^-_1$ + $L^+_2$ & $F^-_1$ + $F^+_2$\\
\hline\\
\end{tabular}
\ \\
\begin{flushleft}
Notes to Table~\ref{tab:wf166}
\end{flushleft}
\begin{enumerate}
\item $z = 0.23\ldots$\ \cite{degty}; the exact value of $z$ is
  meaningless in this table. In the hexagonal unit cell, the Bi atoms
  lie at the Wyckoff positions $6c (00\pm z)$ \cite{degty}. In the
  trigonal system, their positions in the unit cell are $\bm{\rho} =
  \pm (z\bm{T}_1 + z\bm{T}_2 + z\bm{T}_3)$, where the vectors
  $\bm{T}_1, \bm{T}_2,$ and $\bm{T}_3$ denote the basic vectors of the
  trigonal lattice as given, e.g., in Table 3.1 of Ref.~\cite{bc}.
\item The notations of the representations are defined in Table~\ref{tab:rep166}.
\item Assume a closed band of the symmetry in one of the two rows of
  this table to exist in the band structure of Bi--I.  Then the
  Bloch functions of this band can be unitarily transformed into
  Wannier functions that are
\begin{itemize}
\item localized as well as possible; 
\item centered at the Bi atoms; and
\item symmetry-adapted to the space group $R\overline{3}m$ (166)~\cite{theoriewf}.  
\end{itemize}
The entry ``OK'' below the time-inversion operator $K$ indicates that
the Wannier functions may even be chosen symmetry-adapted to the magnetic group
$$  M = R\overline{3}m + K\cdot R\overline{3}m,$$
see Theorem 7 of Ref.~\cite{theoriewf}.\\

However, a closed band (Definition 2 of Ref.~\cite{theoriewf}) with
the symmetry of band 1 or band 2 does not exist in the band structure
of Bi--I (see Fig.~\ref{fig:bs_166}).

\item The bands are determined following Theorem 5 of Ref.\
  \cite{theoriewf}.
\item The Wannier functions at the Bi atoms listed in the upper row
  belong to the representation $\bm{d}_i$ of $C_{3v}$ included below
  the atom. These representations are defined in
  Table~\ref{tab:repBi}.
\item Each row defines one band consisting of two branches, because there
  are two Bi atoms in the unit cell.
\end{enumerate}
\end{table}

\begin{table}
\caption{
Double-valued representations of the superconducting band in the space group
$R\overline{3}m$ of
Bi--I. 
\label{tab:wf166Z}}
\centering
\begin{tabular}[t]{cccccccc}
& Bi($zzz$) & Bi($\overline{z}\overline{z}\overline{z}$) & $K$ & $\Gamma$ & $Z$ & $L$ & $F$\\
\hline
Band 1 & $\bm{d}$ & $\bm{d}$ & OK & $\Gamma^+_4$ + $\Gamma^-_4$ & $Z^+_4$ + $Z^-_4$ & $L^+_3$ + $L^+_4$ + $L^-_3$ + $L^-_4$ & $F^+_3$ + $F^+_4$ + $F^-_3$ + $F^-_4$\\
\hline\\
\end{tabular}\hspace{1cm}
\ \\
\begin{flushleft}
Notes to Table~\ref{tab:wf166Z}
\end{flushleft}
\begin{enumerate}
\item $z = 0.23\ldots$\ \cite{degty}; the exact value of $z$ is
  meaningless in this table. In the hexagonal unit cell, the Bi atoms
  lie at the Wyckoff positions $6c (00\pm z)$ \cite{degty}. In the
  trigonal system, their positions in the unit cell are $\bm{\rho} =
  \pm (z\bm{T}_1 + z\bm{T}_2 + z\bm{T}_3)$, where the vectors
  $\bm{T}_1, \bm{T}_2,$ and $\bm{T}_3$ denote the basic vectors of the
  trigonal lattice as given, e.g., in Table 3.1 of Ref.~\cite{bc}.
\item Assume an isolated band of the symmetry listed in this table to exist in the
  band structure of Bi--I.
  Then the Bloch functions of this band can be unitarily transformed into
  spin dependent Wannier functions that are
\begin{itemize}
\item localized as well as possible; 
\item centered at the Bi atoms; and
\item symmetry-adapted to the space group $R\overline{3}m$ (166)~\cite{theoriewf}.  
\end{itemize}
  The entry ``OK'' below the time-inversion operator $K$ indicates
  that the spin-dependent Wannier functions may even be
  chosen symmetry-adapted to the magnetic group 
  $$  M = R\overline{3}m + K\cdot R\overline{3}m,$$
  see Theorem 10 of
  Ref.~\cite{theoriewf}.
  Hence, the listed band forms a superconducting band, see Definition
  22 of Ref.~\cite{theoriewf}.
\item The listed band is the only superconducting band of Bi--I.  
\item The notations of the double-valued
  representations are (indirectly) defined by Table~\ref{tab:falten166}.
\item Following Theorem 9 of Ref.~\cite{theoriewf}, the superconducting band
  is simply determined from one of the two single-valued bands listed in
  Table~\ref{tab:wf166} by means of Equation (97) of
  Ref.~\cite{theoriewf}. (According to Definition 20 of
  Ref.~\cite{theoriewf}, both single-valued bands in
  Table~\ref{tab:wf166} are affiliated bands of the superconducting
  band.)
\item The superconducting band consists of two branches, because there
  are two Bi atoms in the unit cell.
\item The point group of the positions of the Bi atoms (Definitions 11
  and 12 of Ref.~\cite{theoriewf}) is the group $C_{3v}$.  The Wannier
  functions at the Bi atoms belong to the double-valued representation
  \begin{equation}
    \label{eq:2}
    \bm{d} = \bm{d}_1 \otimes \bm{d}_{1/2} = \bm{d}_2 \otimes \bm{d}_{1/2}
  \end{equation}
  of $C_{3v}$ where $\bm{d}_1$ and $\bm{d}_2$ are defined in
  Table~\ref{tab:repBi} and $\bm{d}_{1/2}$ denotes the two-dimensional
  double-valued representation of $O(3)$ as given, e.g., in Table 6.1
  of Ref.~\cite{bc}.  Note that the two representations $\bm{d}_1
  \otimes \bm{d}_{1/2}$ and $\bm{d}_2 \otimes \bm{d}_{1/2}$ are
  equivalent.
\end{enumerate}
\end{table}

\FloatBarrier

\section{Group-theoretical tables for the cubic space group $Im3m$ (229)
of Bi--V}

\begin{table}[b]
\caption{
  Character tables of the single-valued irreducible representations of the
  space group $Im3m$ = $\Gamma^v_{c}O^{9}_{h}$ of 
  Bi--V. 
  \label{tab:rep229}}
\begin{tabular}[t]{ccccccccccc}
\multicolumn{11}{c}{$\Gamma (000)$, $H (\frac{1}{2}\overline{\frac{1}{2}}\frac{1}{2})$}\\
 & $E$ & $I$ & $\sigma_m$ & $C_{2m}$ & $C^{\pm}_{3j}$ & $S^{\pm}_{6j}$ & $C^{\pm}_{4m}$ & $S^{\pm}_{4m}$ & $C_{2p}$ & $\sigma_{dp}$\\
\hline
$\Gamma^+_1$, $H^+_1$ & 1 & 1 & 1 & 1 & 1 & 1 & 1 & 1 & 1 & 1\\
$\Gamma^+_2$, $H^+_2$ & 1 & 1 & 1 & 1 & 1 & 1 & -1 & -1 & -1 & -1\\
$\Gamma^-_2$, $H^-_2$ & 1 & -1 & -1 & 1 & 1 & -1 & -1 & 1 & -1 & 1\\
$\Gamma^-_1$, $H^-_1$ & 1 & -1 & -1 & 1 & 1 & -1 & 1 & -1 & 1 & -1\\
$\Gamma^+_3$, $H^+_3$ & 2 & 2 & 2 & 2 & -1 & -1 & 0 & 0 & 0 & 0\\
$\Gamma^-_3$, $H^-_3$ & 2 & -2 & -2 & 2 & -1 & 1 & 0 & 0 & 0 & 0\\
$\Gamma^+_4$, $H^+_4$ & 3 & 3 & -1 & -1 & 0 & 0 & 1 & 1 & -1 & -1\\
$\Gamma^+_5$, $H^+_5$ & 3 & 3 & -1 & -1 & 0 & 0 & -1 & -1 & 1 & 1\\
$\Gamma^-_4$, $H^-_4$ & 3 & -3 & 1 & -1 & 0 & 0 & 1 & -1 & -1 & 1\\
$\Gamma^-_5$, $H^-_5$ & 3 & -3 & 1 & -1 & 0 & 0 & -1 & 1 & 1 & -1\\
\hline\\
\end{tabular}\hspace{1cm}
\begin{tabular}[t]{cccccc}
\multicolumn{6}{c}{$P (\frac{1}{4}\frac{1}{4}\frac{1}{4})$}\\
 & $E$ & $C_{2m}$ & $S^{\pm}_{4m}$ & $\sigma_{dp}$ & $C^{\pm}_{3j}$\\
\hline
$P_1$ & 1 & 1 & 1 & 1 & 1\\
$P_2$ & 1 & 1 & -1 & -1 & 1\\
$P_3$ & 2 & 2 & 0 & 0 & -1\\
$P_4$ & 3 & -1 & 1 & -1 & 0\\
$P_5$ & 3 & -1 & -1 & 1 & 0\\
\hline\\
\end{tabular}\hspace{1cm}
\begin{center}
\begin{tabular}[t]{ccccccccc}
\multicolumn{9}{c}{$N (00\frac{1}{2})$}\\
 & $E$ & $C_{2z}$ & $C_{2b}$ & $C_{2a}$ & $I$ & $\sigma_z$ & $\sigma_{db}$ & $\sigma_{da}$\\
\hline
$N^+_1$ & 1 & 1 & 1 & 1 & 1 & 1 & 1 & 1\\
$N^+_2$ & 1 & -1 & 1 & -1 & 1 & -1 & 1 & -1\\
$N^+_3$ & 1 & 1 & -1 & -1 & 1 & 1 & -1 & -1\\
$N^+_4$ & 1 & -1 & -1 & 1 & 1 & -1 & -1 & 1\\
$N^-_1$ & 1 & 1 & 1 & 1 & -1 & -1 & -1 & -1\\
$N^-_2$ & 1 & -1 & 1 & -1 & -1 & 1 & -1 & 1\\
$N^-_3$ & 1 & 1 & -1 & -1 & -1 & -1 & 1 & 1\\
$N^-_4$ & 1 & -1 & -1 & 1 & -1 & 1 & 1 & -1\\
\hline\\
\end{tabular}\hspace{1cm}
\end{center}
\ \\
\begin{flushleft}
Notes to Table~\ref{tab:rep229}
\end{flushleft}
\begin{enumerate}
\item $m = x, y, z;\quad p = a, b, c, d, e, f;\quad j = 1, 2, 3, 4.$
\item The symmetry elements are labeled in the Sch\"onflies notation
  as illustrated, e.g., in Table 1.2 of Ref.~\cite{bc}.  
\item The character tables are determined from Table 5.7 of
  Ref.~\protect\cite{bc}.
\item The notations of the points of symmetry follow Fig. 3.15 of 
Ref.~\cite{bc}. 
\end{enumerate}
\end{table}

\begin{table}
\caption{
    Compatibility relations between the single-valued (upper row) and
    double-valued (lower row) representations of the space group
    $Im3m$. 
\label{tab:falten229}}
\centering
\begin{tabular}[t]{cccccccccc}
\multicolumn{10}{c}{$\Gamma (000)$, $H (\frac{1}{2}\overline{\frac{1}{2}}\frac{1}{2})$}\\
\hline
$R^+_1$ & $R^+_2$ & $R^-_2$ & $R^-_1$ & $R^+_3$ & $R^-_3$ & $R^+_4$ & $R^+_5$ & $R^-_4$ & $R^-_5$\\
$R^+_6$ & $R^+_7$ & $R^-_7$ & $R^-_6$ & $R^+_8$ & $R^-_8$ & $R^+_6$ + $R^+_8$ & $R^+_7$ + $R^+_8$ & $R^-_6$ + $R^-_8$ & $R^-_7$ + $R^-_8$\\
\hline\\
\end{tabular}\hspace{1cm}
\begin{tabular}[t]{ccccc}
\multicolumn{5}{c}{$P (\frac{1}{4}\frac{1}{4}\frac{1}{4})$}\\
\hline
$P_1$ & $P_2$ & $P_3$ & $P_4$ & $P_5$\\
$P_6$ & $P_7$ & $P_8$ & $P_6$ + $P_8$ & $P_7$ + $P_8$\\
\hline\\
\end{tabular}\hspace{1cm}
\begin{tabular}[t]{cccccccc}
\multicolumn{8}{c}{$N (00\frac{1}{2})$}\\
\hline
$N^+_1$ & $N^+_2$ & $N^+_3$ & $N^+_4$ & $N^-_1$ & $N^-_2$ & $N^-_3$ & $N^-_4$\\
$N^+_5$ & $N^+_5$ & $N^+_5$ & $N^+_5$ & $N^-_5$ & $N^-_5$ & $N^-_5$ & $N^-_5$\\
\hline\\
\end{tabular}\hspace{1cm}
\ \\
\begin{flushleft}
Notes to Table~\ref{tab:falten229}
\end{flushleft}
\begin{enumerate}
\item In the table for $\Gamma$ and $H$, the letter $R$ stands for the letter denoting the point of
  symmetry.  For example, at point $H$ the representations $R^+_1, R^+_2, \ldots$
  stand for $H^+_1, H^+_2, \ldots$ .
\item Each column lists the double-valued representation $R_i\times
  {\bm d}_{1/2}$
  below the single-valued representation $R_i$, where ${\bm d}_{1/2}$ denotes the
  two-dimensional double-valued representation of the three-dimensional
  rotation group $O(3)$ given, e.g., in Table 6.1 of Ref.~\cite{bc}.
\item The single-valued representations are defined in
  Table~\ref{tab:rep229}. 
\item The notations of double-valued representations follow strictly
  Table 6.13 (and Table 6.14) of Ref.~\cite{bc}. In this paper the
  double-valued representations are not explicitly given but are
  sufficiently defined by this table.
\end{enumerate}
\end{table}
\begin{table}
\caption{
Single-valued representations of the space group $Im3m$ of all the energy bands of
Bi--V with symmetry-adapted and optimally  
localized usual (i.e., spin-independent) Wannier functions centered at the Bi atoms. 
\label{tab:wf229}}
\centering
\begin{tabular}[t]{ccccccc}
 & Bi($000$) & $K$ & $\Gamma$ & $H$ & $P$ & $N$\\
\hline
Band 1 & $\Gamma^+_1$ & OK & $\Gamma^+_1$ & $H^+_1$ & $P_1$ & $N^+_1$\\
Band 2 & $\Gamma^+_2$ & OK & $\Gamma^+_2$ & $H^+_2$ & $P_2$ & $N^+_3$\\
Band 3 & $\Gamma^-_2$ & OK & $\Gamma^-_2$ & $H^-_2$ & $P_1$ & $N^-_3$\\
Band 4 & $\Gamma^-_1$ & OK & $\Gamma^-_1$ & $H^-_1$ & $P_2$ & $N^-_1$\\
\hline\\
\end{tabular}
\ \\
\begin{flushleft}
Notes to Table~\ref{tab:wf229}
\end{flushleft}
\begin{enumerate}
\item The notations of the representations are defined in Table~\ref{tab:rep229}.
\item Assume a closed band of the symmetry in any row of this table to exist in the
  band structure of Bi--V.
  Then the Bloch functions of this band can be unitarily transformed into
  Wannier functions that are
\begin{itemize}
\item localized as well as possible; 
\item centered at the Bi atoms; and
\item symmetry-adapted to the space group $Im3m$ (229)~\cite{theoriewf}.  
\end{itemize}
The entry ``OK'' below the time-inversion operator $K$ indicates that
the Wannier functions may even be chosen symmetry-adapted to the magnetic group
$$  M = Im3m + K\cdot Im3m,$$
see Theorem 7 of Ref.~\cite{theoriewf}.\\

However, a closed band (Definition 2 of Ref.~\cite{theoriewf}) with
the symmetry of the bands in this table does not exist in the band
structure of Bi--V (see Fig.~\ref{fig:bs_229}).

\item The bands are determined following Theorem 5 of Ref.\
  \cite{theoriewf}.
\item The point group of the positions of the Bi atoms (Definitions 11
  and 12 of Ref.~\cite{theoriewf}) is the full cubic point group $O_h$.
The Wannier functions at the Bi atoms belong to the
  representations of $O_h$ listed in the second column. These
  representations are defined in Table~\ref{tab:rep229}.
\end{enumerate}
\end{table}


\begin{table}
\caption{
Double-valued representations of the space group $Im3m$ of all the 
energy bands of Bi--V with symmetry-adapted and 
optimally localized spin-dependent Wannier functions centered at 
the Bi atoms. 
\label{tab:wf229Z}}
\centering
\begin{tabular}[t]{ccccccc}
 & Bi($000$) & $K$ & $\Gamma$ & $H$ & $P$ & $N$\\
\hline
Band 1 & $\Gamma^+_1 \otimes {\bm d}_{1/2} = \Gamma^+_6$ & OK & $\Gamma^+_6$ & $H^+_6$ & $P_6$ & $N^+_5$\\
Band 2 & $\Gamma^+_2 \otimes {\bm d}_{1/2} = \Gamma^+_7$ & OK & $\Gamma^+_7$ & $H^+_7$ & $P_7$ & $N^+_5$\\
Band 3 & $\Gamma^-_2 \otimes {\bm d}_{1/2} = \Gamma^-_7$ & OK & $\Gamma^-_7$ & $H^-_7$ & $P_6$ & $N^-_5$\\
Band 4 & $\Gamma^-_1 \otimes {\bm d}_{1/2} = \Gamma^-_6$ & OK & $\Gamma^-_6$ & $H^-_6$ & $P_7$ & $N^-_5$\\
\hline\\
\end{tabular}
\ \\
\begin{flushleft}
Notes to Table~\ref{tab:wf229Z}
\end{flushleft}
\begin{enumerate}
\item Assume an isolated band of the symmetry listed in any row of this
  table to exist in the band structure of Bi--V.
  Then the Bloch functions of this band can be unitarily transformed
  into spin-dependent Wannier functions that are
\begin{itemize}
\item localized as well as possible; 
\item centered at the Bi atoms; and
\item symmetry-adapted to the space group $Im3m$ (229)~\cite{theoriewf}.  
\end{itemize}
  The entry ``OK'' below the time-inversion operator $K$ indicates
  that the spin dependent Wannier functions may even be
  chosen symmetry-adapted to the magnetic group 
  $$  M = Im3m + K\cdot Im3m,$$
  see Theorem 10 of
  Ref.~\cite{theoriewf}.
  Hence, all the listed bands forms superconducting bands, see Definition
  22 of Ref.~\cite{theoriewf}.
\item The notations of the double-valued
  representations are (indirectly) defined in Table~\ref{tab:falten229}.
\item Following Theorem 9 of Ref.~\cite{theoriewf}, the
  superconducting bands are simply determined from the single-valued
  bands listed in Table~\ref{tab:wf229} by means of Equation (97) of
  Ref.~\cite{theoriewf}. (According to Definition 20 of
  Ref.~\cite{theoriewf}, each single-valued band in
  Table~\ref{tab:wf229} is an affiliated band of one of the
  superconducting bands.)
\item The superconducting bands consists of one branch each, because
  there is one Bi atom in the unit cell.
\item The point group of the positions of the Bi atoms (Definitions 11
  and 12 of Ref.~\cite{theoriewf}) is the full cubic point group
  $O_h$.  The Wannier functions at the Bi atoms belong to the
  double-valued representations of $O_h$ listed in the second column,
  where the single-valued representations $\Gamma^{\pm}_1$ and
  $\Gamma^{\pm}_2$ are defined by Table~\ref{tab:rep229}, and
  $\bm{d}_{1/2}$ denotes the two-dimensional double-valued
  representation of $O(3)$ as given, e.g., in Table 6.1 of
  Ref.~\cite{bc}.
\end{enumerate}
\end{table}

\FloatBarrier


\end{document}